\documentclass{aa}
\usepackage[]{natbib,amsmath,amssymb}
\usepackage{graphicx}
\bibpunct{(}{)}{;}{a}{}{,}
\begin{document}
\title{Mass-sheet degeneracy: Fundamental limit on the cluster mass 
  reconstruction from statistical (weak) lensing}
\titlerunning{Mass-sheet degeneracy in statistical (weak) lensing cluster mass 
  reconstructions}
\author{M. Brada\v{c} \inst{1,2} \and M. Lombardi \inst{1,3}
  \and P. Schneider \inst{1}
} 
\offprints{Maru\v{s}a Brada\v{c}}
\mail{marusa@astro.uni-bonn.de}
\institute{Institut f\"ur Astrophysik und Extraterrestrische
  Forschung, Auf dem H\"ugel 71, D-53121 Bonn, Germany
  \and Max-Planck-Institut f\"ur Radioastronomie, Auf dem
  H\"ugel 69, D-53121 Bonn, Germany
  \and European Southern Observatory, Karl-Schwarzschild-Str. 2,
  D-85748 Garching bei M\"unchen, Germany 
}
\date{Received November 25, 2003; accepted May 11, 2004}
%
%

\abstract{Weak gravitational lensing is considered to be one of the
  most powerful tools to study the mass and the mass distribution of
  galaxy clusters.  However, weak lensing mass reconstructions are
  plagued by the so-called mass-sheet degeneracy~--~the surface mass
  density $\kappa$ of the cluster can be determined only up to a degeneracy
  transformation $\kappa \to \kappa' = \lambda \kappa + (1 -
  \lambda)$, where $\lambda$ is an arbitrary
  constant.  This transformation fundamentally limits the accuracy of
  cluster mass determinations if no further assumptions are made.  We
  describe here a method to break the mass-sheet degeneracy in weak
  lensing mass maps using distortion and redshift information of
  background galaxies and illustrate this by two simple toy models.  
Compared to other techniques proposed in the
  past, it does not rely on any assumptions on cluster potential;
  it can be easily applied to non-parametric mass-reconstructions and
  no assumptions on boundary conditions have to be made. In addition it
  does not make use of weakly constrained information (such as the
  source number counts, used in the magnification effect).  
Our simulations show that \textit{we are effectively able to break the
    mass-sheet degeneracy for supercritical lenses}, but
that for undercritical lenses the mass-sheet degeneracy is very
  difficult to be broken, even under idealised conditions.
   }

\maketitle
%
%

\def\diff{\mathrm{d}}

\def\eck#1{\left\lbrack #1 \right\rbrack}
\def\eckk#1{\bigl[ #1 \bigr]}
\def\rund#1{\left( #1 \right)}
\def\abs#1{\left\vert #1 \right\vert}
\def\wave#1{\left\lbrace #1 \right\rbrace}
\def\ave#1{\left\langle #1 \right\rangle}

\def\vec#1{%
  \if\alpha#1\mathchoice%
    {\mbox{\boldmath$\displaystyle#1$}}%
    {\mbox{\boldmath$\textstyle#1$}}%
    {\mbox{\boldmath$\scriptstyle#1$}}%
    {\mbox{\boldmath$\scriptscriptstyle#1$}}%
  \else
    \textbf{\textit{#1}}%
  \fi}

%
%
\section{Introduction}
\label{sec:introduction}

Galaxy clusters have been the focus of a very intense ongoing research
in the past years.  They are the most massive bound structures in the
Universe; moreover, their large dynamical time scale allows them to retain
the information about their formation history.  The mass and the mass
distribution of galaxy clusters is particularly important for
cosmological studies and therefore provides a critical 
test of the Cold Dark Matter
(CDM) paradigm.

Weak gravitational lensing techniques are a unique tool to measure
the projected mass of a cluster.  This effect has long been recognized
and measured (see \citealp{tyson90}), however it was only after the
pioneering work of \citeauthor{kaiser93} (\citeyear{kaiser93}; see
e.g. \citealp{clowe01, clowe02} for further applications) that the
field began to flourish.  Since then several other mass reconstruction
techniques and applications of weak lensing have been developed
\citep[see e.g.][]{ksb95, lombardibertin99, seitz97, bridle98,
  marshall02, hoekstra98}.

Unfortunately, all these methods suffer from the fact that the
projected surface mass density $\kappa$ can be determined only up to a
degeneracy transformation $\kappa \to \kappa' = \lambda \kappa + (1-\lambda)$, where $\lambda$ is an
arbitrary constant. This invariance (first recognized by
\citealt{falco85} in the context of strong lensing and by \citet{seitz95}
in the context of weak lensing) leaves the relation between the intrinsic and
observed ellipticity unchanged. Therefore it can not be broken by
using only measurements of the distortion of the background sources,
if these all lie at the same distance from the observer and 
fundamentally limits the accuracy of cluster mass
determinations if no further assumptions are made.

A naive solution to this problem is to constrain $\lambda$ by making simple
assumptions about $\kappa$.  For example, one can assume that the surface mass
density is decreasing with distance from the centre, implying $\lambda >
0$.  In addition, $\kappa$ is likely to be non-negative, and so one can
obtain an upper limit on $\lambda$ (for $\kappa < 1$).

More quantitatively, with the use of wide field cameras one might try
to assume that $\kappa \simeq 0$ at the boundary of the field, far
away from the
cluster center.  However, if we consider for example a $M_\mathrm{vir}
= 10^{15} M_{\odot}$ cluster at redshift $z = 0.2$, we expect from N-body
simulations to have a projected dimensionless density of about $\kappa \simeq
0.005$ at $15 \mbox{ arcmin}$ from the cluster center (Douglas Clowe,
private communication).  Hence, even with the use of a $30 \times 30 \mbox{
  arcmin}$ camera we expect to underestimate the virial mass of such a
cluster by $\sim 20 \%$.

 Another approach to determine the mass of a cluster from weak
  lensing data is to fit the shear signal with a parametric model
  (such as SIS or NFW). However, as shown in \citet{clowe01, clowe02},
  it is extremely difficult to distinguish SIS from NFW at high
  significance, and thus the mass estimates become difficult. 
  They conclude that one of the main sources of uncertainty in 
determining the cluster mass 
  and profile determination is the mass-sheet degeneracy.

Unfortunately, the mass-sheet degeneracy can not be lifted by using
the shapes of the background sources alone \citep{seitz97}, and so one
needs to make use of additional information.  One promising solution
to this problem is provided by the so-called magnification effect
\citep[see][]{broadhurst95}.  Indeed, the local number counts of
background sources is directly related to the magnification of the
lens, which to first order is given by $1 + 2 \kappa$.  As a result, a
careful estimate of the local density of background sources can lead
to a direct measurement of the projected density $\kappa$ (see
\citealp{fort97, taylor98} for reported detections of this effect). In
addition, other methods to measure cluster masses via magnification
effects have been proposed (e.g. \citealp{bartelmann95} suggested to use
the sizes of galaxies at a given surface brightness
as a measure of the local lens magnification).  All these methods,
however, require a fine calibration of external parameters, such as
the number counts of unlensed sources, which is very difficult to
obtain with desired accuracy \citep{schneider00}.

In this work we will therefore focus on the possible methods to break
the mass-sheet degeneracy by using distortion and redshift information
of background sources.  \citet{seitz97} already considered this
problem, and they showed that the mass-sheet degeneracy
can be \textit{weakly\/} broken provided that the probability
distribution of the redshifts for the observed sources is known with
good accuracy. Recently, it has been shown that photometric redshifts based on
accurate multi-band photometry can be extremely accurate 
\citep[see e.g.][]{benitez00}.  
Hence, it is sensible to assume that the \textit{individual\/}
redshifts of the background sources can be known in weak lensing
studies. As we will show later, this in principle allows one to break
the mass-sheet degeneracy.

This paper is organised as follows. In Sect.~\ref{sc:statlens} we give the
basic lens equations, introduce the notation we use, and introduce the
problem of mass-sheet degeneracy. In 
Sect.~\ref{sec:clust-mass-reconstr} we focus on the cluster mass
reconstruction technique and we describe the problem of
mass-sheet degeneracy in more detail in
Sect.~\ref{sec:mass-sheet-degen-1}. In Sect.~\ref{sec:simulated-data} 
we describe the simulated data we are using and present the results. 
We conclude in Sect.~\ref{sec:conclusions}. 

\section{Statistical lensing preliminaries}
\label{sc:statlens}

\subsection{Notation and basic lens relationships}
\label{sc:lens}

In this paper we use the standard weak lensing notation as described
in \citet{bartelmann00}.  We consider a lens with projected surface
mass density $\Sigma(\vec \theta)$, where $\vec \theta$ denotes the position in the lens
plane.  For a source at redshift $z$ and a lens at redshift
$z_\mathrm{d}$ we define the critical surface mass 
density $\Sigma_\mathrm{cr}(z)$ as
\begin{equation}
  \label{eq:1}
  \Sigma_\mathrm{cr} = \frac{c^2}{4\pi G} \,
  \frac{D_\mathrm{s}}{D_\mathrm{d} D_\mathrm{ds}} \; ,
\end{equation}
where $D$ represents the angular diameter distances; 
$D_\mathrm{s} = D(0,z)$, $D_\mathrm{d} = D(0,z_\mathrm{d})$, and
$D_\mathrm{ds} = D(z_\mathrm{d}, z)$ are the angular diameter
distances between the observer and the source, the observer and the
lens, and the lens and the source, respectively.

Since we will work with sources at different redshifts, we factorize
the redshift dependence of the lens convergence $\kappa$ and of the shear
$\gamma$ as
\begin{align}
  \label{eq:2}
  \kappa(\vec\theta, z) & {} = Z(z) \kappa(\vec\theta) \; , &
  \gamma(\vec\theta, z) & {} = Z(z) \gamma(\vec\theta) \; , &
\end{align}
where $Z(z)$ is the so-called ``cosmological weight'' function: 
\begin{equation}
  \label{eq:419}
  Z(z) \equiv \frac{\lim_{z\to\infty}\Sigma_\mathrm{cr}(z_\mathrm{d},z)}%
  {\Sigma_\mathrm{cr}(z_\mathrm{d},z)} \, \mathrm{H}(z-z_\mathrm{d}) \; .
\end{equation}
The Heaviside step function $\mathrm{H}(z-z_\mathrm{d})$ accounts for
the fact that the sources that are closer to the observer than the
deflector are not lensed.  Note that, as suggested by its name, $Z(z)$
is cosmology dependent.  For an Einstein-de Sitter cosmology it becomes
\begin{equation}
  \label{eq:3}
  Z(z) =\frac{\sqrt{1+z} - \sqrt{1+z_\mathrm{d}}}{\sqrt{1+z} - 1} \, \mathrm{H}(z-z_\mathrm{d}) \; .
\end{equation}
In \citet{lombardibertin99} cosmological weights were calculated for
different cosmologies.  The authors have shown that the differences
between Einstein-de Sitter and the nowadays assumed standard cosmology
(i.e.\ $\Omega_{\rm m} = 0.3$, $\Omega_\Lambda = 0.7$) are not significant for the
purpose of cluster-mass reconstructions.  Therefore we will from now on
use Einstein-de Sitter cosmology. Similarly to $\kappa$ and $\gamma$, 
we define the redshift-dependent
reduced shear $g(\vec\theta, z)$ as
\begin{equation}
  \label{eq:4}
  g(\vec\theta, z) = \frac{Z(z) \gamma(\vec\theta)}{1 - Z(z) \kappa(\vec\theta)} \; .
\end{equation}

Throughout this paper we use the complex ellipticity $\epsilon$
defined in terms of the second brightness moments $Q_{ij}$ as
\begin{equation}
  \label{eq:5}
  \epsilon \equiv \frac{Q_{11} - Q_{22} + 2\mathrm{i}Q_{12}}%
  {Q_{11} + Q_{22} + 2(Q_{11}Q_{22} - Q_{12}^2)^{1/2}}\;.
\end{equation}
The transformation between the source ellipticity $ \epsilon^{\rm s}$
and image ellipticity $\epsilon$ is given as a function of reduced
shear $g(\vec \theta, z)$ \citep[see][]{seitz97},
\begin{equation}
  \label{eq:6}
  \epsilon^{\rm s} = 
  \begin{cases}
    \dfrac{\epsilon - g(\vec \theta, z)}{1 - g^*(\vec \theta, z)\epsilon} & \text{for $\bigl\lvert
      g(\vec\theta, z) \bigr\rvert \leq 1 \; ,$} \\[1.5em]
    \dfrac{1 - g(\vec \theta, z)\epsilon^*}{\epsilon^* - g^*(\vec \theta, z)} & \text{for $\bigl\lvert
      g(\vec\theta, z) \bigr\rvert > 1 \; .$}
  \end{cases}
\end{equation}
The inverse transformation is simply given by swapping $\epsilon^\mathrm{s}$
with $\epsilon$ and by substituting $-g$ for $g$.

Under the assumption that the intrinsic ellipticity distribution is
isotropic, $\bigl\langle \epsilon^{\rm s} \bigr\rangle = 0$, the expectation
value for the image ellipticity at redshift $z$ becomes
\begin{equation}
  \label{eq:7}
  \bigl\langle \epsilon(z) \bigr\rangle = \begin{cases}
    g(\vec\theta, z) & \text{if $\bigl\lvert g(\vec\theta, z) \bigr\rvert < 1 \;
    ,$} \\[1em]
    \frac{\displaystyle 1}{\displaystyle g^*(\vec\theta, z)}
    & \text{otherwise$\; .$}
  \end{cases}
\end{equation}
In the weak lensing approximation which we define by  $\kappa \ll
  1$, $\abs{\gamma} \ll 1$  (thus $\abs{g} \ll 1$) the
  expectation value is given by 
$\bigl\langle \epsilon(z) \bigr\rangle = \gamma(\vec\theta,z)$.

\subsection{The problem of the mass sheet degeneracy}
\label{sc:mass-sheet}
In the simple case of  background sources all having the {\it same 
redshift}, the
mass-sheet degeneracy can be understood just using the above
equations.  Indeed, consider for a moment 
the transformation of the potential $\psi$
\begin{equation}
  \label{eq:8a}
  \psi(\vec\theta, z) \to \psi'(\vec\theta, z) = 
  \frac{1-\lambda}{2} \vec\theta^2 +  \lambda\psi(\vec\theta, z)\; ,
\end{equation}
where $\lambda$ is an arbitrary constant. 
$\kappa$ and $\gamma$ are related to the potential $\psi$ through its
second partial derivatives (denoted by subscript), namely 
\begin{equation}
  \label{eq:8aa}
\kappa = \frac{1}{2}(\psi_{,11}+\psi_{,22})\; ,\;  \gamma_1 =
  \frac{1}{2}(\psi_{,11}-\psi_{,22}) \; , \; \gamma_2 =
  \psi_{,12}\; .
\end{equation}
From \eqref{eq:8a} it follows that $\kappa$ transforms as 
\begin{equation}
  \label{eq:8}
  \kappa(\vec\theta, z) \to \kappa'(\vec\theta, z) = 
  \lambda \kappa(\vec\theta, z) + (1 - \lambda) \; ,
\end{equation}
and  similarly the shear changes as
$\gamma(\vec\theta, z)$ to $\lambda \gamma(\vec\theta, z)$. Therefore
the reduced shear $g(\vec\theta, z)$ remains invariant.

\citet{seitz97} have shown that in the case of a known {\it redshift
  distribution}, a similar form of the mass-sheet
degeneracy holds to a very good approximation  for non-critical 
clusters, i.e.\ for clusters
with $|g(\vec \theta, z)| \leq 1$ for all source redshifts $z$.  In
  such a case the standard weak-lensing mass reconstruction is affected by the
degeneracy
\begin{equation}
  \label{eq:9}
  \kappa \to \kappa' \simeq \lambda\kappa+ \frac{(1-\lambda)
    \:\ave{Z(z)}}{\ave{Z^2(z)}} \; ,
\end{equation}
where $\bigl\langle Z^n(z) \bigr\rangle$ denotes the $n$-th order moment of the
distribution of cosmological weights.  As a result,
\textit{standard\/} weak-lensing reconstructions are still affected by
the mass-sheet degeneracy even for sources at different redshifts;
moreover, simulations show that the degeneracy is hardly broken even
for lenses close to critical.

In this paper, we use the information of {\it individual redshifts} of
background sources to break this degeneracy. As an illustration of the
effect, suppose that half of the
background sources are located at a known redshift $z^{(1)}$, and the
other half at another known redshift $z^{(2)}$.  Then, the weak lensing
reconstructions based on the two populations will provide two
different mass maps, $\kappa'(\vec\theta, z^{(1)})$ and
$\kappa'(\vec\theta, z^{(2)})$, leading to two different forms of the 
mass-sheet
degeneracy. In other words, the two mass reconstructions $(i=1,2)$ are
given by
\begin{equation}
  \label{eq:901}
  \kappa'(\vec\theta, z^{(i)}) = \lambda^{(i)} \kappa_{\rm
  t}(\vec\theta, z^{(i)}) + \bigl(1 - \lambda^{(i)} \bigr)
\end{equation}
where we have denoted $\kappa_{\rm t}(\vec\theta,z^{(i)})$ 
the true projected $\kappa$
of the lens at the angular position $\vec\theta$ 
for sources at redshift $z^{(i)}$. Since 
the transformation \eqref{eq:901} holds for any
$\vec\theta$, we have  a system of equations
to be solved for $\lambda^{(1)}$ and $\lambda^{(2)}$. 
The relation between $\kappa_{\rm
  t}(\vec\theta,z^{(1)})$ and  $\kappa_{\rm
  t}(\vec\theta,z^{(2)})$ is known, namely from Eq.~\eqref{eq:2} follows 
\begin{equation}
  \label{eq:901a}
\kappa_{\rm t}(\vec\theta,z^{(1)}) Z(z^{(2)}) = \kappa_{\rm
  t}(\vec\theta,z^{(2)}) Z(z^{(1)}).
\end{equation}
Suppose one  measures
both  $\kappa'(\vec\theta, z^{(i)})$ at $N$ different positions 
$\vec\theta_{j}$, this gives
us a system of  $2N$ equations to be solved for $\lambda^{(i)}$ and 
$\kappa_{\rm t}(\vec\theta_{j})$.  The mass-sheet degeneracy is
therefore at least in theory lifted.

It is interesting to observe that
this argument only applies to relatively ``strong'' lenses.  Indeed,
for ``weak'' lenses, i.e.\ lenses for which we can use a first order
approximation in $\kappa$ and $\gamma$, the expectation value of
measured image ellipticities is $\bigl\langle
\epsilon(z) \bigr\rangle = \gamma(\vec\theta, z)$. In such case  
the degeneracy of the form
\begin{equation}
  \label{eq:902}
 \psi(\vec\theta, z) \to \psi'(\vec\theta, z) = 
  \frac{1-\lambda}{2} \vec\theta^2 +  \psi(\vec\theta, z)
\end{equation}
leaves the observable $\gamma(\vec\theta, z)$ unchanged. 
As a result, the method described above cannot be
used to break the mass-sheet degeneracy for these lenses. Only when
the $(1-Z(z)\kappa)$ term in the reduced shear becomes important and
$g(\vec\theta,z)$ can be distinguished from $\gamma(\vec\theta, z)$ in
the (noisy) data, we will be able to make unbiased cluster mass 
reconstructions.

The aperture mass measures \citep{kaiser95,schneider96} 
are very convenient for measuring the weak
lensing signal. A simple form of such measures is the so-called
$\zeta$ statistics, 
\begin{equation}
 \label{eq:902b}
\zeta(\vec\theta;\vartheta_1,\vartheta_2) \equiv
  \bar\kappa(\vec\theta;\vartheta_1)
  -\bar\kappa(\vec\theta;\vartheta_1,\vartheta_2)\;.
\end{equation}
which gives the difference between the mean surface mass densities
in a circle of radius $\vartheta_1$ at position $\vec\theta$ and in an
annulus between the radii  $\vartheta_1$ and  $\vartheta_2$. It
measures the mass contrast, and thus can not be used to determine the
absolute mass scale.
Therefore despite the fact that it is not affected by the mass-sheet
degeneracy transformation of the form \eqref{eq:902}, it will not be
discussed further in this work.

In conclusion, using individual redshifts of background sources
 provides the most direct method to break the mass-sheet degeneracy. 
The simple argument sketched above for two redshift planes 
will therefore be considered in more detail
in this paper and generalised to background galaxies at different
redshifts. In the following we will use mock catalogues of image
ellipticities, generated under optimistic assumptions. Such an approach will
allow us to investigate the minimum lensing strength needed to be able to lift
the degeneracy in the forthcoming observations.  
As we will see below, this straightforward method can be
successfully used to break the mass-sheet degeneracy in statistical
 lensing when applied in the strong
lensing regime.

\section{Cluster mass reconstruction from image distortion}
\label{sec:clust-mass-reconstr}

\subsection{Maximum-likelihood approach}
\label{sec:maxim-likel-appr}

In order to further investigate the effect of using individual
redshifts in the cluster mass reconstructions, 
it is useful to adopt a Bayesian approach and to write the
likelihood function of a given observed configuration of galaxy
ellipticities.

Let us call $p_{\epsilon^\mathrm{s}}(\epsilon^\mathrm{s})$ the probability
distribution of \textit{intrinsic\/} source ellipticities. For
simplicity and better understanding we will here (and throughout this
paper) assume the following truncated Gaussian distribution 
  \begin{equation}
    \label{eq:20}
    p_{\epsilon^\mathrm{s}}(\epsilon^\mathrm{s}) = \frac{1}{2 \pi \sigma^2
      \bigl[ 1 - \exp \bigl( - 1 / 2 \sigma^2 \bigr) \bigr]} 
    \exp \bigl( - \bigl\lvert \epsilon^\mathrm{s} \bigr\rvert^2 / 2 \sigma^2 \bigr)
    \; .
  \end{equation}
Note however that the results in this section are independent of the
particular choice for $p_{\epsilon^\mathrm{s}}$.
The
\textit{observed\/} distribution $p_{\epsilon}(\epsilon \mid g)$ is related to the
intrinsic one by \citep[see][]{geiger98}
\begin{equation}
  \label{eq:10}
  p_{\epsilon}(\epsilon \mid g) = p_{\epsilon^\mathrm{s}} \bigl( \epsilon^\mathrm{s}(\epsilon \mid g) \bigr)
  \left| \frac{\mathrm{d}^2 \epsilon^\mathrm{s}}{\mathrm{d}^2 \epsilon} \right| 
  (\epsilon \mid g) \; .
\end{equation}
The Jacobian determinant from the previous equation is given by
\begin{equation}
  \label{eq:11}
  \left| \frac{\mathrm{d}^2 \epsilon^\mathrm{s}}{\mathrm{d}^2 \epsilon} \right| (\epsilon \mid
  g) = 
  \begin{cases}
    \dfrac{(|g|^2-1)^2}{|\epsilon\,g^{\star}-1|^4} & \text{for $|g| \leq 1 \; ,$} \\[1.5em]
    \dfrac{(|g|^2-1)^2}{|\epsilon-g|^4} & \text{for $|g| > 1 \; .$}
  \end{cases}
\end{equation}
In the weak lensing limit, $\abs{g} \ll 1$, the
Jacobian determinant is simply unity.

In general, the \textit{measured\/} ellipticity $\epsilon^\mathrm{m}$
will differ from $\epsilon$ because of measurement errors:
\begin{equation}
  \label{eq:12}
  \epsilon^\mathrm{m} = \epsilon + \epsilon^\mathrm{err}\; .
\end{equation}
The error $\epsilon^\mathrm{err}$ is a random variable whose distribution
depends on the details of the ellipticity measurement algorithm.  In
the following, for simplicity (and throughout this paper), 
we will assume that $\epsilon^{\rm err}$ is distributed according 
to a Gaussian distribution
with dispersion $\sigma_\mathrm{err}$. The actual shape
measurement which corrects for PSF smearing (see e.g. \citealp{ksb95})
can yield measured ellipticities with $\abs{\epsilon^\mathrm{m}} > 1$.

In our case, we write the probability distribution for the measured 
ellipticities as
\begin{equation}
  \label{eq:13}
  p_{\epsilon^\mathrm{m}}(\epsilon^\mathrm{m} \mid g)=
  \int p_{\epsilon}(\epsilon \mid g) \, p_{\epsilon^\mathrm{err}}(\epsilon - \epsilon^\mathrm{m}) \diff \epsilon\;,
\end{equation}
where $p_{\epsilon^\mathrm{err}}(\epsilon^\mathrm{err})$ is the
probability distribution of measurement errors. This convolution
takes into account the above mentioned measurement errors and therefore we
do not need to discard galaxies with  $\abs{\epsilon^\mathrm{m}} >
1$.

The likelihood function $\mathcal{L}$ is the product of probability
densities for observed image ellipticities $\epsilon_i^{\rm m}$, and depends
on the model parameters $\pi$ through the reduced shear at the image
positions $\{ g_i \}$:
\begin{equation}
  \label{eq:14}
  \mathcal{L}(\pi) = \prod_{i=1}^{N_\mathrm{g}}
  p_{\epsilon^\mathrm{m}}(\epsilon_i^\mathrm{m} \mid g_i) \; .
\end{equation}
Here $N_\mathrm{g}$ is the number of observed galaxies with measured
ellipticities $\epsilon_i^\mathrm{m}$.  It is more convenient to deal
with the log-likelihood function, defined as
\begin{equation}
  \label{eq:15}
  l(\pi) \equiv - \ln \mathcal{L}(\pi) = -\sum_{i=1}^{N_\mathrm{g}} \ln p_{\epsilon}(\epsilon_i \mid
  g) \; .
\end{equation} 
By minimising $l(\pi)$ we obtain the most likely parameters $\pi_\mathrm{
  max}$ given the observations.

For each of the $N_\mathrm{g}$ galaxies we
  need to calculate the probability distribution  
$p_{\epsilon^\mathrm{m}}(\epsilon^\mathrm{m} \mid g)$ [cf.\
  Eq.~\eqref{eq:13}] of measured image ellipticity 
$\epsilon_i^\mathrm{m}$
  given the reduced shear $g$. The evaluation of the full likelihood 
function is
  therefore non-trivial, since 
  $N_\mathrm{g}$ integrals need to be calculated. In simple words, we
  are dealing with two Gaussian probability 
distributions for source ellipticities $\epsilon^\mathrm{s}$ and the measurement error 
$\epsilon^\mathrm{err}$. However, the final measured
errors for observed image ellipticities $\epsilon^\mathrm{m}$ is not 
distributed according to a Gaussian in general, 
since the source ellipticities are first being lensed and
  only then the measurement errors are added (i.e. the Jacobian
  determinant in \eqref{eq:10} is not unity).

If the reduced shear is small (i.e. for \textit{undercritical\/}
  lenses), this problem can be solved by including the measurement
  errors in the \textit{source\/} probability distribution
  $p_{\epsilon^\mathrm{ s}}$: in other words, if $|g| \ll 1$, the
  Jacobian determinant is unity and $p_{\epsilon}(\epsilon \mid g) =
  p_{\epsilon^\mathrm{s}} \bigl( \epsilon^\mathrm{s}(\epsilon \mid g)
  \bigr)$. We can therefore interchange the convolution with
  $p_{\epsilon^\mathrm{err}}$ appearing in Eq.~\eqref{eq:13} with the
  lensing transformation of Eq.~\eqref{eq:10}.  The calculations are
  then trivial, since $p_{\epsilon^\mathrm{s}}$ and
  $p_{\epsilon^\mathrm{err}}$ can be taken to be Gaussians with
  dispersions $\sigma_{\epsilon^\mathrm{s}}^2$ and
  $\sigma_\mathrm{err}^2$ respectively; hence, we can just use for
  $p_{\epsilon^\mathrm{s}}$ a Gaussian with dispersion
  $\sigma_{\epsilon^\mathrm{s}}^2 + \sigma_\mathrm{err}^2$.

This is, however, not the case for critical lenses.  As we show
latter, the approximate form for the measured ellipticities [cf.\
Eq.~\eqref{eq:10}] gives biased results for lenses with a large
fraction of background sources having $|g| \simeq 1$.  For the same
reason, we should not discard galaxies with $\abs{\epsilon^\mathrm{m}}
> 1$ without properly accounting for their removal in
Eq.~\eqref{eq:13}. Therefore, unless otherwise stated, we will use
the probability distribution for measured ellipticity given in
Eq.~\eqref{eq:13}. It still does not account for the measurement
errors on the redshifts of background sources, however these are less
important.

\subsection{$\chi^2$ approach}
\label{sec:chi-appr}
Finally, we note an alternative to the maximum-likelihood approach,
namely a chi-square fitting. In our case, the $\chi^2$ function can
be written as
\begin{equation}
  \label{eq:16}
  \chi^2(\pi) = \sum_{i=1}^{N_\mathrm{g}} \frac{\abs{\epsilon_i -
      \langle \epsilon \rangle}^2}{\sigma_i^2} \; ,
\end{equation}
where 
\begin{equation}
  \label{eq:17}
  \sigma_i^2 = \rund{1- \bigl\lvert \langle \epsilon \rangle
  \bigr\rvert^2}^2 \sigma^2_{\epsilon^{\rm s}}
  + \sigma^2_{\rm err}\; .
\end{equation}
Unfortunately, simulations show that such an approach does not give an
unbiased result for critical lenses.  This is due to two main reasons.
First, the expression \eqref{eq:17} is only an approximation for the
true variance, that contains higher (even) order moments of the
distribution of $\epsilon^\mathrm{s}$ \citep{marco_thesis}.  Second
and more important, ellipticity errors are not normally distributed
for lenses with large reduced shears, $|g| \simeq 1$ (see
\citealp{geiger98}, Fig.~5).  In the maximum likelihood approach, on
the other hand, the real distribution of errors is properly accounted
for and in the asymptotic limit the method will therefore give an
unbiased result.  Hence, in the following we will only discuss the
likelihood approach.

\section{The mass-sheet degeneracy} 
\label{sec:mass-sheet-degen-1}

From the discussion above, it is apparent that only galaxies with $|g|
\simeq 1$ contribute significantly to the removal of the mass-sheet
degeneracy for two reasons:
\begin{itemize}
\item The dispersion on the observed ellipticities is proportional to
  $\bigl( 1- \bigl\lvert \langle \epsilon \rangle^2 \bigr\rvert
  \bigr)^2$ [cf.\ Eq.~\eqref{eq:17}], and so galaxies with $|g| \simeq
  1$ provide more
  accurate estimates on the local reduced shear;
\item Recalling the argument discussed after Eq.~\eqref{eq:901},
  galaxies with $|g| \ll 1$ present a simple mass-sheet degeneracy [cf.\ 
  Eq.~\eqref{eq:902}] which cannot be broken using the method
  described here. Therefore one needs the information from  
$|g| \simeq 1$ galaxies to break the degeneracy.
\end{itemize}
It is important to realize that the likelihood function in the way we
wrote it, already takes into account the first point mentioned above.
As a result, galaxies with $|g| \simeq 1$ are effectively more
important for the purpose of breaking the mass-sheet degeneracy, while
those with $|g| \ll 1$ do not contribute at all (as sketched in
Sect.~\ref{sc:mass-sheet}).  It is sensible, therefore, to take into
account this point when writing the approximate form of the mass-sheet
degeneracy presented in Eq.~\eqref{eq:9}. In other words, we
assume the following transformation of $\bigl\langle Z^n (z)
\bigr\rangle$
\begin{equation}
  \label{eq:18}
   \bigl\langle Z^n(z) \bigr\rangle \to \bigl\langle Z^n(z) \bigr\rangle' = \frac{\sum_{i=1}^{N_\mathrm{g}}
    w_i Z_i^n}%
  {\sum_{i=1}^{N_\mathrm{g}} w_i} \; ,
\end{equation}
where $w_i$ are the weight factors and for $w_i = \mbox{const.}$ we recover
the results of \citet{seitz97}. For giving more weights to galaxies with $|g|
\simeq 1$, we will however use  $w_i = 1 / \sigma_i$, defined in 
\eqref{eq:17}. As we mentioned before this is only an approximation
for the true dispersion, however as we will show latter it is adequate
to describe the approximate mass-sheet degeneracy behaviour.

\section{Simulated data}
\label{sec:simulated-data}

\begin{figure*}[ht!]
\begin{minipage}{17cm}
\begin{minipage}{8.5cm}
\begin{center}
\includegraphics[width=0.9\textwidth]{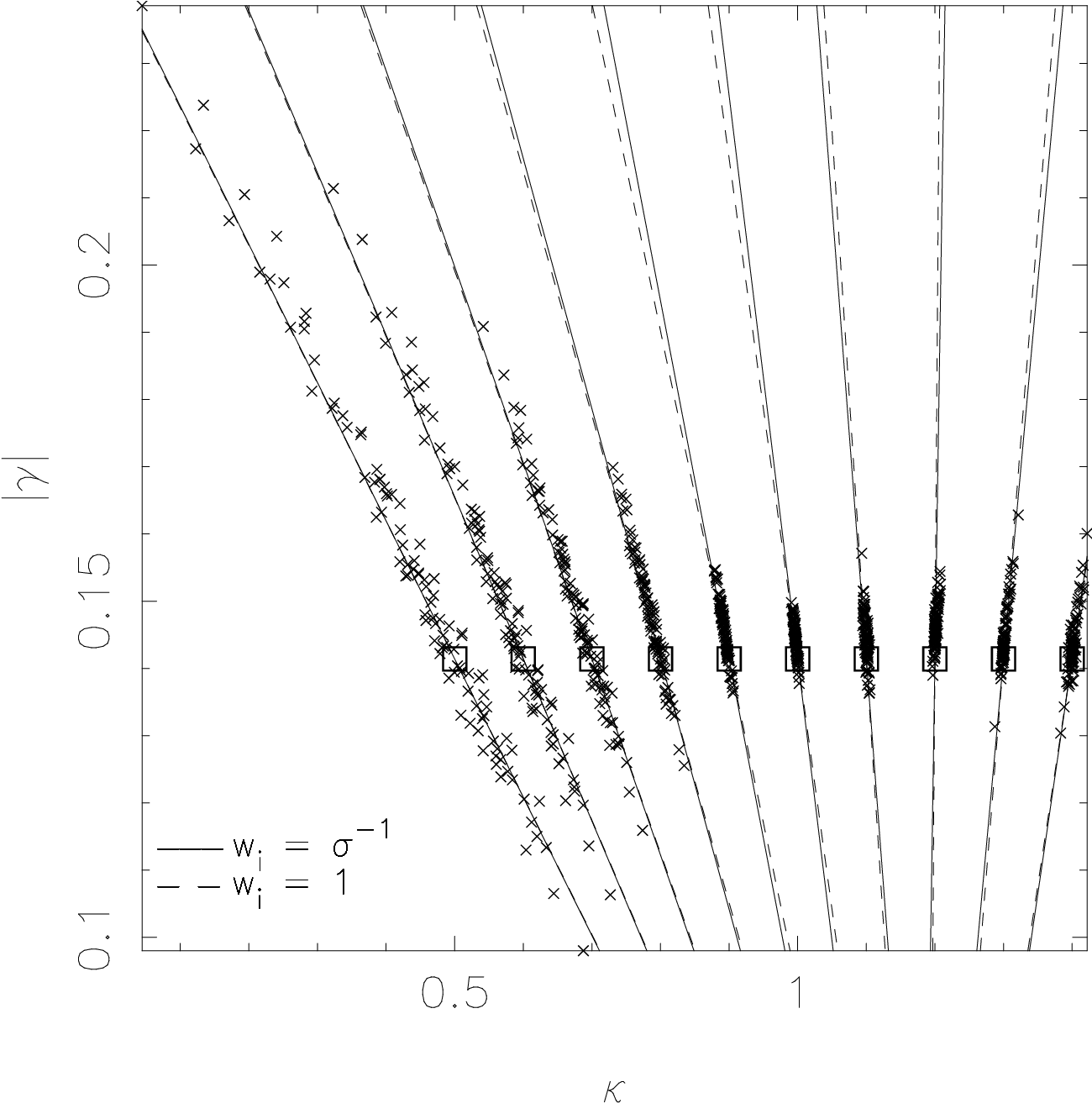}
\end{center}
\end{minipage}
\begin{minipage}{8.5cm}
\begin{center}
\includegraphics[width=0.9\textwidth]{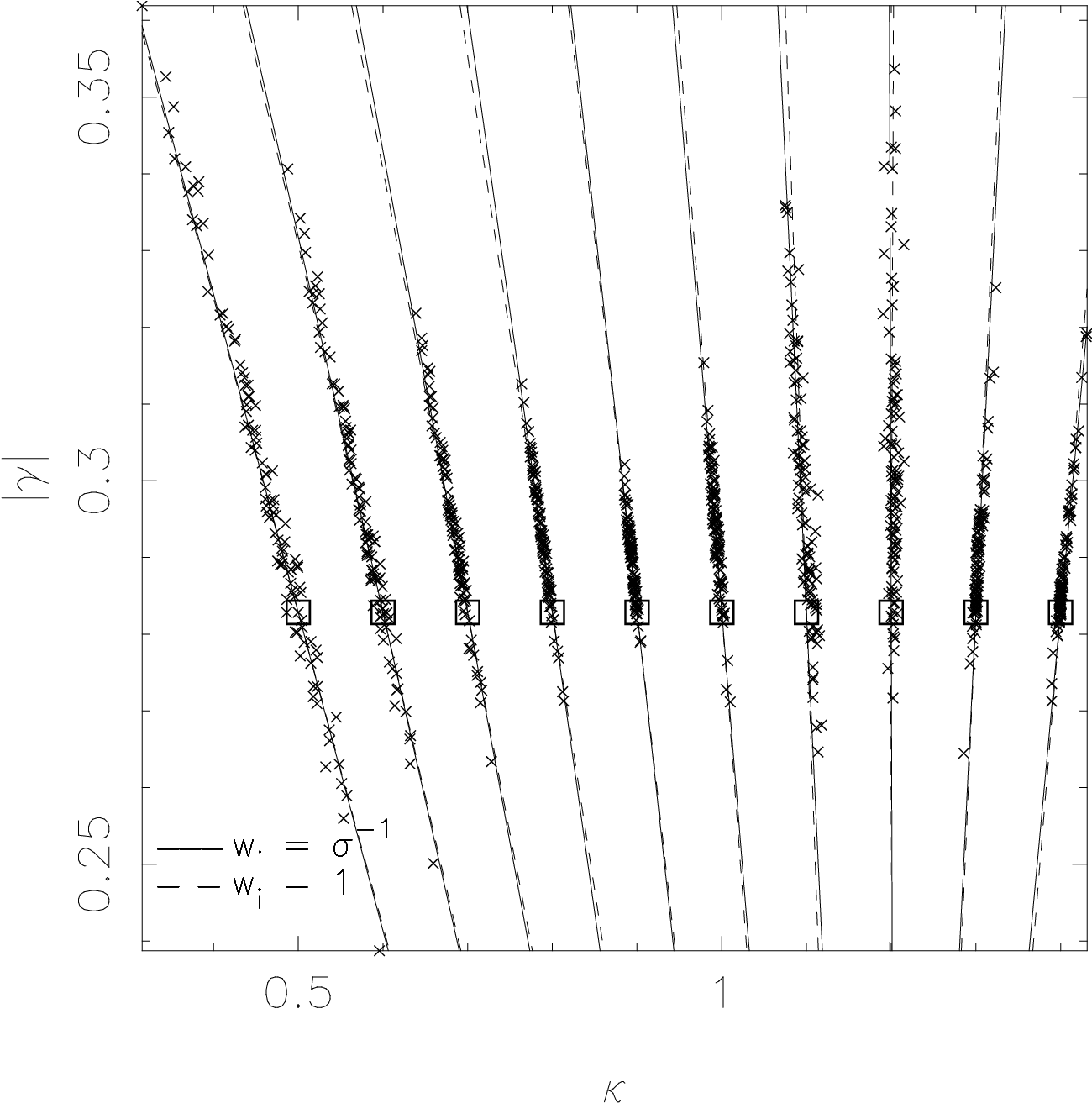}
\end{center}
\end{minipage}
\end{minipage}
\
\vspace{0.5cm}
\
\begin{minipage}{17cm}
\begin{minipage}{8.5cm}
\begin{center}
\includegraphics[width=0.9\textwidth]{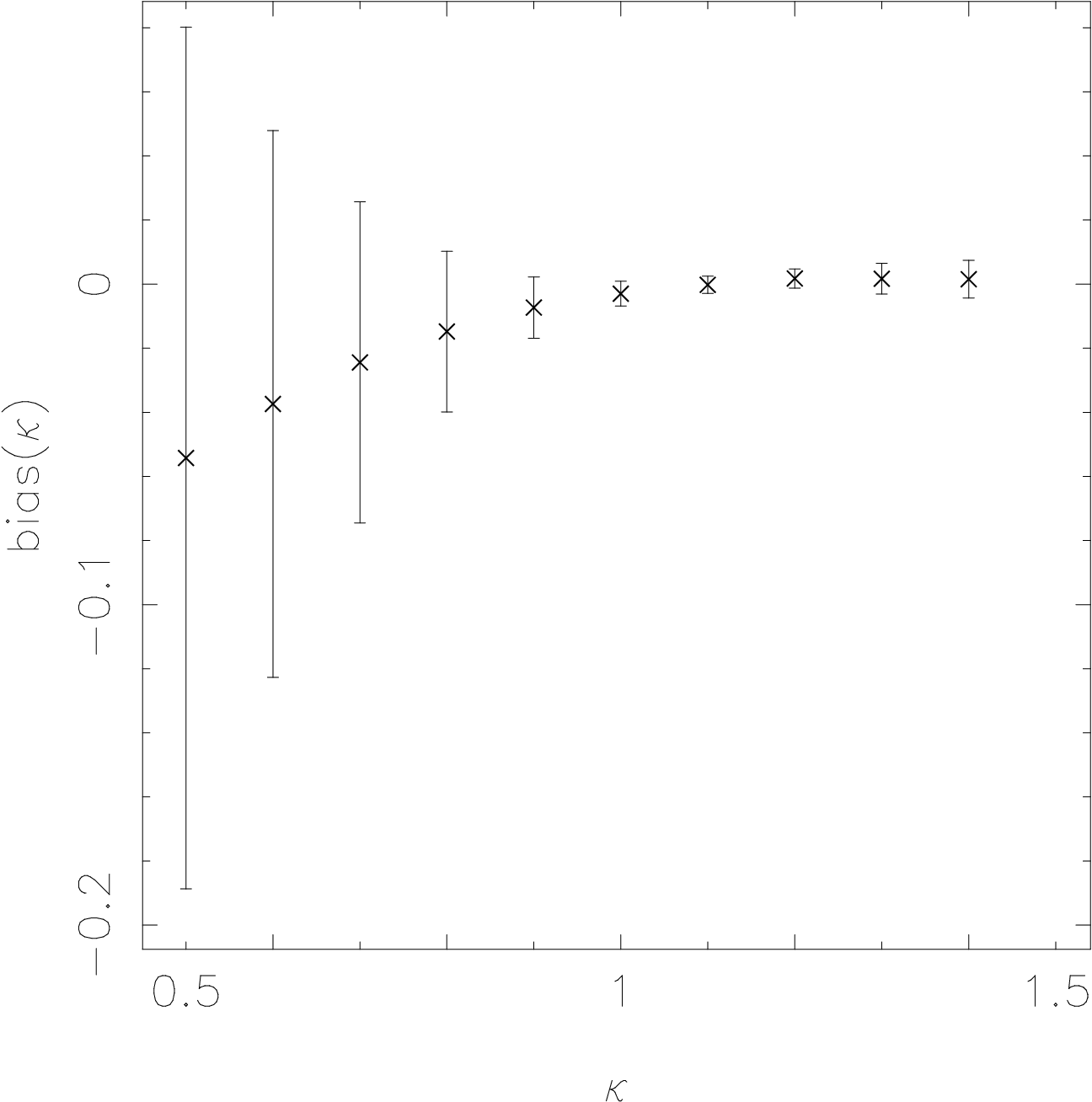}
\centerline{\bf (a)}
\end{center}
\end{minipage}
\begin{minipage}{8.5cm}
\begin{center}
\includegraphics[width=0.9\textwidth]{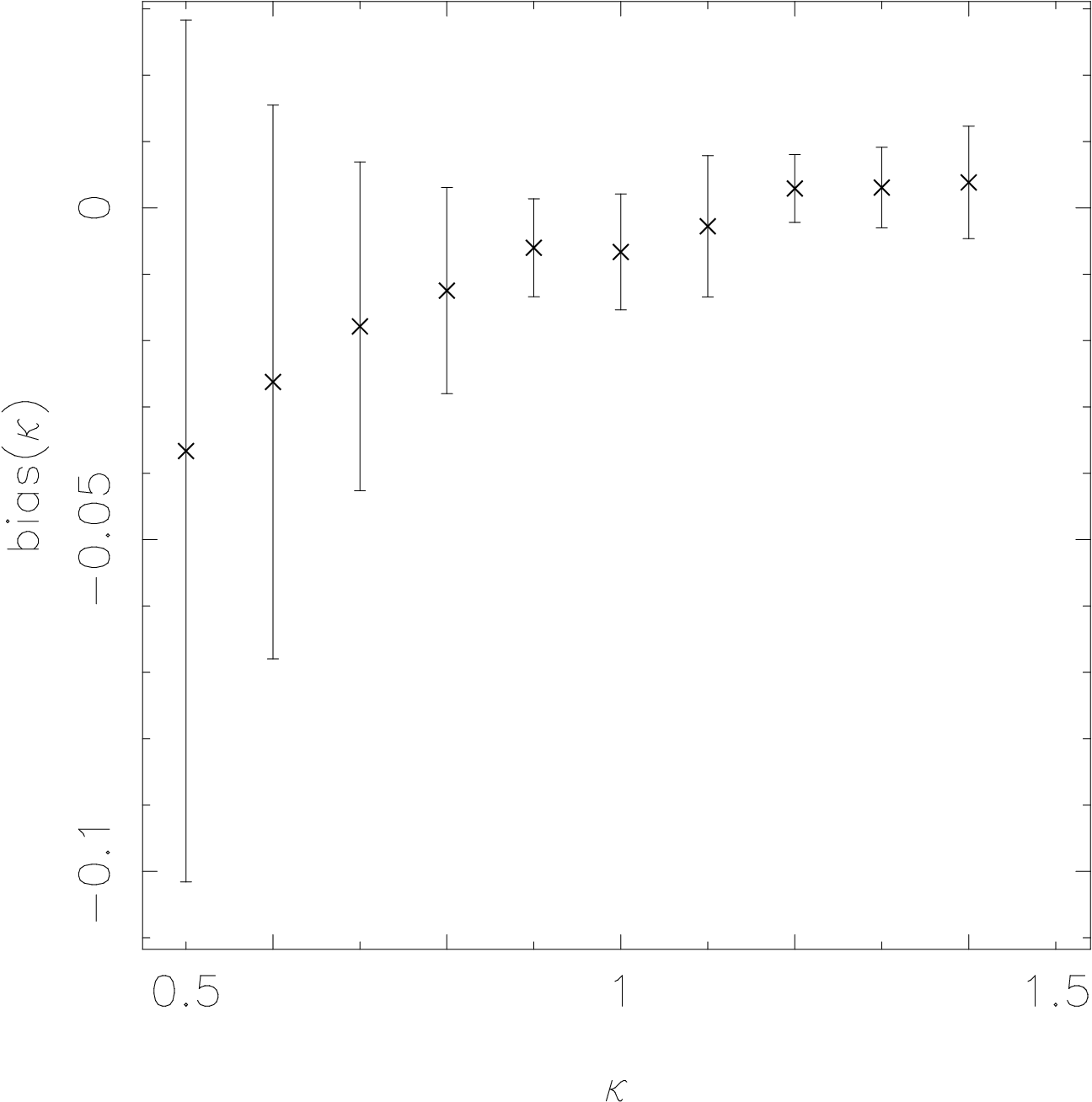}
\centerline{\bf (b)}
\end{center}
\end{minipage}
\end{minipage}
\caption{Top: Recovered parameter values (crosses) as a result of
  minimising the log-likelihood function \eqref{eq:15}.  For each of
  the ten sets of parameters (panel {\bf(a)} $\gamma_1 = \gamma_2 =
  0.1$, panel {\bf(b)} $\gamma_1 = \gamma_2 = 0.2$) -- denoted by
  squares -- 100 mock catalogues were created using model Family I and
  the same family was used to fit the data. We use 3 free parameters
  ($\kappa$, $\gamma_1$, and $\gamma_2$) for fitting, here we plot
  $\kappa$ and $\abs{\gamma}$.  Lines correspond to the expected
  mass-sheet degeneracy calculated using the weighting scheme given by
  $w_i = 1 / \sigma_i$ (solid lines) and $w_i = {\rm const.}$ (dashed
  lines) in \eqref{eq:18}. Bottom: Bias and variance for the recovered
  values of $\kappa$ for the data described above.}
\label{fig:fitconstlens}
\end{figure*}

\begin{figure*}[ht!]
\begin{minipage}{17cm}
\begin{minipage}{8.5cm}
\begin{center}
\includegraphics[width=0.9\textwidth]{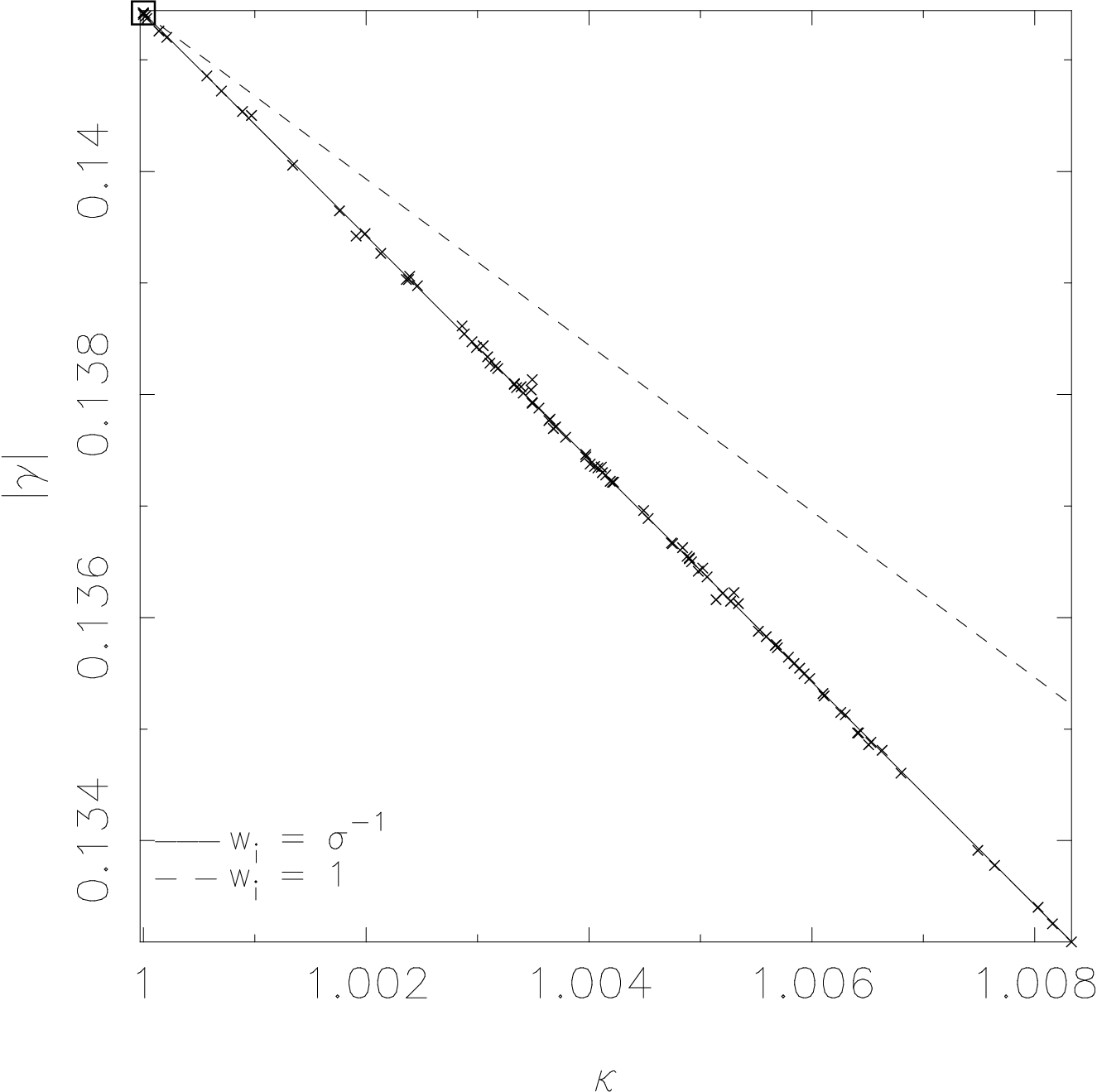}
\centerline{\bf (a)}
\end{center}
\end{minipage}
\begin{minipage}{8.5cm}
\begin{center}
\includegraphics[width=0.9\textwidth]{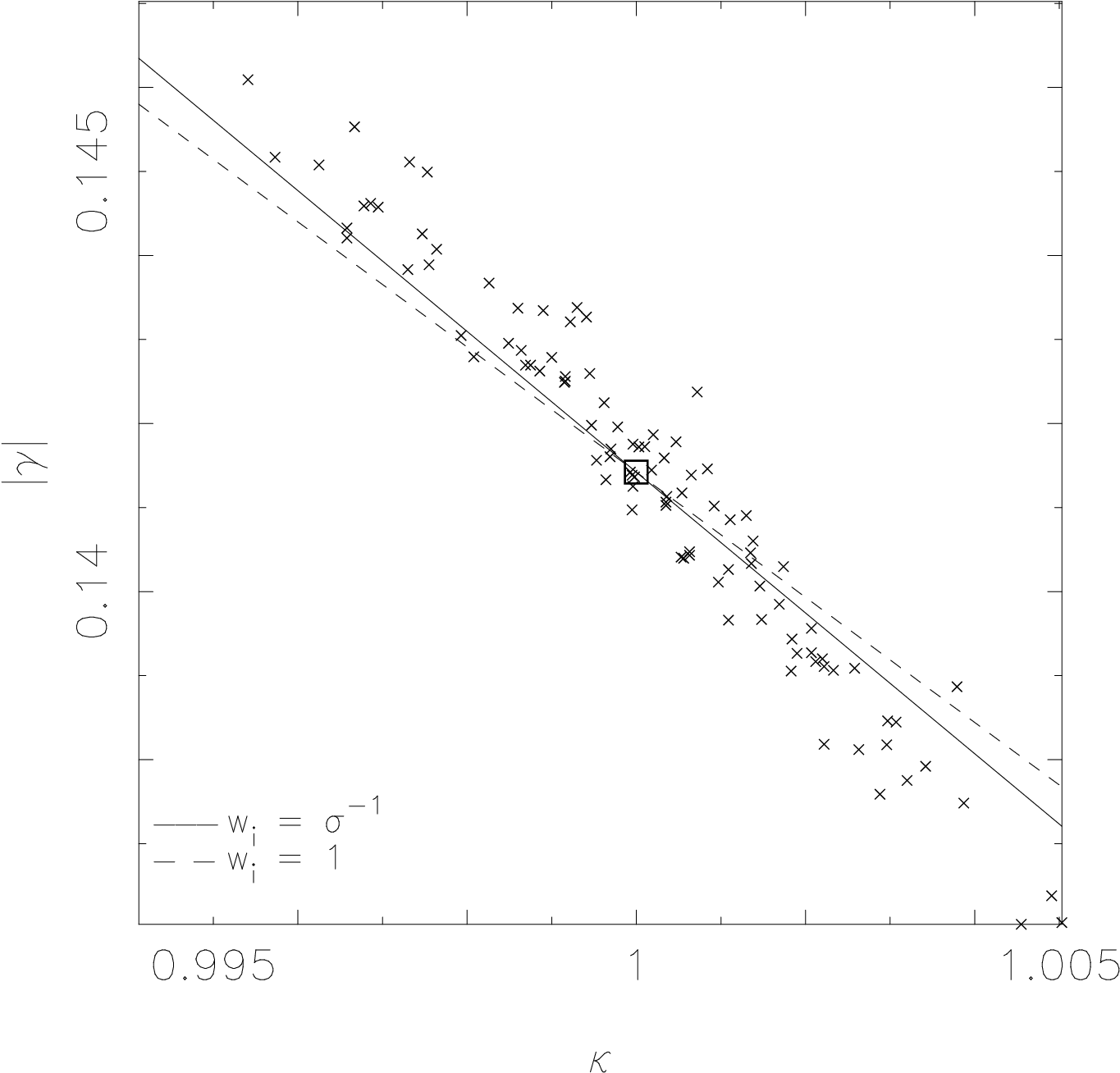}
\centerline{\bf (b)}
\end{center}
\end{minipage}
\end{minipage}
\caption{Recovered parameter values (crosses) as a result of
  minimising the log-likelihood function Eq.~\eqref{eq:15}. Panel
  \textbf{(a)} shows results obtained using $p_{\epsilon}(\epsilon
  \mid g)$ -- Eq.~\eqref{eq:10} -- for the observed probability
  distribution for ellipticities, while for the panel \textbf{(b)} we
  use (as in the rest of this paper)
  $p_{\epsilon^\mathrm{m}}(\epsilon^\mathrm{m} \mid g)$ --
  Eq.~\eqref{eq:13}.  For the parameter set $\kappa = 1.0$, $\gamma_1
  = \gamma_2 = 0.1$ -- denoted by a square -- 100 mock catalogues were
  created using model Family I and the same family was used to fit the
  data. Redshift errors were {\it not} added. We use 3 free parameters
  ($\kappa$, $\gamma_1$, and $\gamma_2$) for fitting, here we plot
  $\kappa$ and $\abs{\gamma}$. Lines correspond to the expected
  mass-sheet degeneracy calculated using Eq.~\eqref{eq:18}. We used
  the weighting scheme given by $w_i = {\rm const.}$ for the dashed
  lines. For the solid lines $w_i = 1 / \sigma_i$ was used, however
  $\sigma_\mathrm{err} = 0$ was employed for panel \textbf{(a)} only.
  This is due to the fact that we are using $p_{\epsilon}(\epsilon
  \mid g)$ in \textbf{(a)} for the observed probability distribution
  for ellipticities, rather than
  $p_{\epsilon^\mathrm{m}}(\epsilon^\mathrm{m} \mid g)$.}
\label{fig:fit_pde}
\end{figure*}

In order to test whether we can break the mass-sheet degeneracy by
using redshift information on the background sources, we performed two
simple tests on simulated data.

We considered first the most simple model possible, a constant sheet
of mass with external shear (Family~I).  More precisely, we set
$\kappa(\vec \theta) = C_{1}$, $\gamma_1(\vec \theta) = C_{2}$, and $\gamma_2(\vec \theta) = C_{3}$,
with $C_{i}$ constants.  Thus, in this model the convergence and shear
for each galaxy depend only on the redshift of the source and not on
its position.  Although, clearly, this model is completely
unrealistic, it is very useful as a test for our method.  Indeed, it is
reasonable to assume that if we are not able to break the mass-sheet
degeneracy in this simple and favorable situation, it is unlikely that
we will be able to break it in more realistic and complicated cases.
This simple model can also be used as an indicator of how strong the
lens should be in order to obtain a reliable estimate for the average
surface mass density $\kappa$ across the field. 

The second family we used is a non-singular model that approximates an
isothermal sphere for large distances \citep[see][]{sc92} in which we
allow for a constant sheet in surface mass density (hereafter
Family~II).  The dimensionless surface mass density is given by
\begin{equation}
  \label{eq:19}
  \kappa ( \theta / \theta_\mathrm{c} ) = \kappa_0 \frac{1 + \theta^2 /\rund{2 \theta_\mathrm{c}^2}}{\bigl( 1 + \theta^2 / \theta_\mathrm{c}^2 \bigr)^{3/2}} + \kappa_1 \; ,
\end{equation}
where $\kappa_0$, $\kappa_1$ are dimensionless constants and $\theta_\mathrm{c}$ is
the core radius.

We used the following recipe for generating mock catalogues of
background sources:
\begin{enumerate}
\item 2000 galaxies were drawn randomly across the field of $6 \times 6
  \mbox{ arcmin}^2$, thus giving a density of $55 \mbox{ galaxies
    arcmin}^{-2}$.
\item The intrinsic ellipticities $\epsilon^\mathrm{s}$ were drawn
  according to \eqref{eq:20}
  characterized by $\sigma = \sigma_{\epsilon^\mathrm{s}} = 0.15$. 
\item We draw the redshifts of the background sources, following
  \citet{brainerd96} from a gamma distribution
  \begin{equation} 
    p_\mathrm{z}(z) = \frac{z^2}{2\:z_0^3} \exp\rund{-z/z_0}\; ,
    \label{eq:z}
  \end{equation}
  with $z_0 = 2 / 3$; the mean redshift is $\langle z \rangle = 3 z_0
  = 2$, and the mode is $z_\mathrm{mode} = 2 z_0 = 4 / 3$.  The
  corresponding cosmological weights were evaluated assuming an
  Einstein-de~Sitter cosmology. We put the lens at a redshift $z_{\rm
  d} = 0.2$.  For the purpose of mass-sheet degeneracy breaking one
  might think it would be of an advantage to have a higher redshift
  lens, e.g. $z_{\rm d} = 0.4$.  In such a case most of the galaxies
  would lie at the steep part of the function $Z(z)$ and therefore we
  would have a higher scatter of $Z_i$ values. However, this effect
  compensated with the fact, that the average reduced shear is lower
  for a lens at higher redshift. The effect of the number change of
  background sources further favours the low redshift cluster, however
  not significantly.
\item For each galaxy, we evaluated the local shear according to the
  lens model, its position and redshift; then we
  lensed the galaxy ellipticities.  Note that the positions of the
  source galaxies were not transformed, i.e.\ we neglected here the
  magnification effect of the lens.
\item The measurement error $\epsilon^{\rm err}$ on the observed
  ellipticities was drawn from the distribution \eqref{eq:20} with $\sigma =
  \sigma_\mathrm{err} = 0.1$ and added to the lensed ellipticities. 
\item In most cases we considered measurement errors on the redshifts
  of the galaxies to simulate the use of photometric redshifts.  These
  errors were drawn from a Gaussian distribution with $\sigma_\mathrm{zerr}
  = 0.06 \rund{1+z_i}$ \citep[see][]{benitez00}; in adding the errors
  we ensured that the resulting redshifts are  always positive.
\end{enumerate}
In generating the mock catalogues we have tried to simulate an ideal
  case, since we are trying to answer the question when (and if at
  all) the mass-sheet degeneracy can be broken in statistical lensing
  mass reconstructions.  From a comparison between independent
  observations of the same galaxies in COMBO-17 survey
  $\sigma_\mathrm{err} \simeq 0.1 - 0.15$ (for each component of the
  ellipticity) can be estimated \citep{kleinheinrich03}. The typical
  values of $\sigma_{\epsilon^\mathrm{s}}$ are optimistically $\simeq
  0.2$.  However even a less optimistic case
  ($\sigma_{\epsilon^\mathrm{s}} = 0.3$, $\sigma_\mathrm{err} = 0.15$)
  does not change our conclusions.

\subsection{Results of the model fitting to the simulated data}
\label{sec:results-model-fitt}

For each set of true model parameters $\pi_\mathrm{t}$ we generated 100
mock catalogues. Using these, we searched for the most likely
parameters $\pi_\mathrm{max}$ by minimising the log-likelihood function
\eqref{eq:15}. 
We performed the minimisation with \texttt{C-minuit}
\citep{james75}, a routine which is a part of the CERN Program
Library.  \texttt{C-minuit} is designed to minimise a multi-parameter
function and analyse its shape around the minimum. It simultaneously
make use of the gradient as well as downhill simplex method.

Despite the goodness of the routine, one has to be very careful when
performing the minimisation, as the function \eqref{eq:15} has
logarithmic singularities for $|g| = 1$ when $\sigma_{\rm err} = 0$ is
assumed.  If, for a particular
parameter set, a background galaxy happens to have $|g| \simeq 1$, the
log-likelihood function \eqref{eq:15} diverges and the minimisation
procedure has difficulties to ``climb'' over such region, possibly
leading to secondary minima. 

In the case of  $\sigma_{\rm err} \neq 0$ this is less likely to
happen, as the convolution with $p_{\epsilon^\mathrm{err}}$ in 
\eqref{eq:13} is effectively smoothing  the probability
distribution. Unfortunately the integration in \eqref{eq:13} has to be
performed numerically. The noise resulting from numerical calculations is
therefore present 
in the log-likelihood function and the minimisation routine is not able
to search for a minimum, since the function is  not smooth. 
On top of that, the integrations are CPU
consuming, one needs to perform them for every galaxy separately. In
order to avoid these problems, we note that function
$p_{\epsilon^\mathrm{m}}$ \eqref{eq:13} depends on three parameters
(without a loss of generality we can assume $g$ to be real and
accordingly transform $\epsilon^\mathrm{m}$). We therefore 
evaluate $p_{\epsilon^\mathrm{m}}$ on a
three-dimensional grid (with
GNU Scientific Library 
{\tt qags} routine, {\tt http://www.gnu.org/software/gsl}) and use
tri-linear interpolation to evaluate the log-likelihood.

Finally, we stress that the model fitted to the data belonged, for all
simulations, to the same family as the original model used to generate
the galaxy catalog.  Although unrealistic, this assumption allows us
to directly compare the deduced parameters with the original ones, and
thus simplifies the evaluation and interpretation of the simulation
results. The free parameters of both models are
$\pi={\kappa,\gamma_1,\gamma_2}$ for Family I,
$\pi={\kappa_0,\kappa_1}$ for Family II. These correspond to the mass
sheet degeneracy transformation, for the case where redshift
information is not known; $\kappa(\vec\theta) \to \kappa'(\vec\theta)
= \lambda \kappa(\vec\theta, z) + (1 - \lambda)$ and
$\gamma(\vec\theta) \to \gamma'(\vec\theta) = \lambda
\gamma(\vec\theta)$. By investigating the distribution of best fit
parameters $\pi$ for both models, we can answer the question of how
strong the lens need to be in order to lift the above mentioned
degeneracy under idealised conditions.

\subsubsection{Family I}
\begin{figure*}[ht]
\begin{minipage}{17cm}
\begin{minipage}{8.5cm}
\begin{center}
\includegraphics[width=0.8\textwidth]{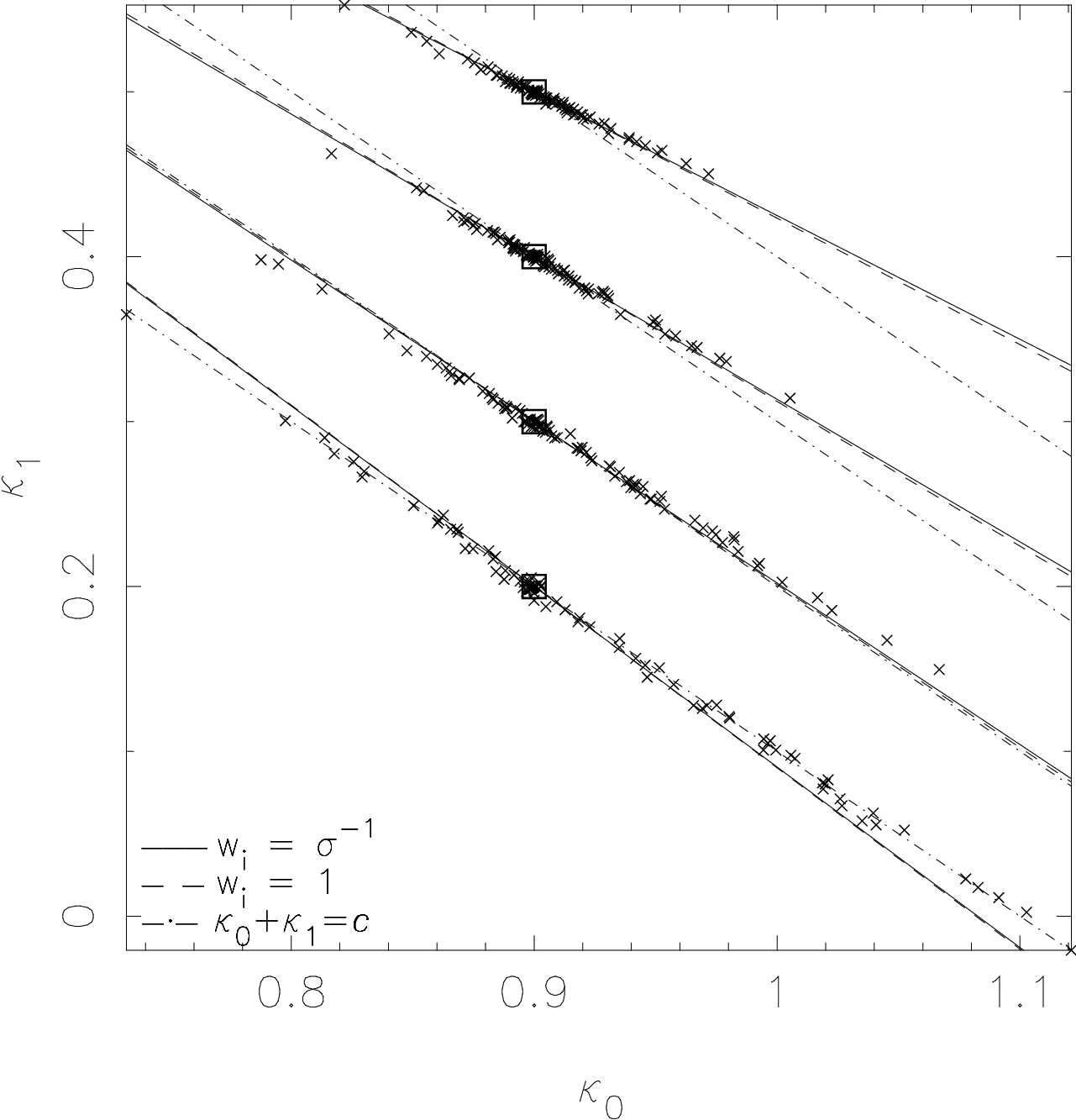}
\centerline{\bf (a)}
\end{center}
\end{minipage}
\begin{minipage}{8.5cm}
\begin{center}
\includegraphics[width=0.8\textwidth]{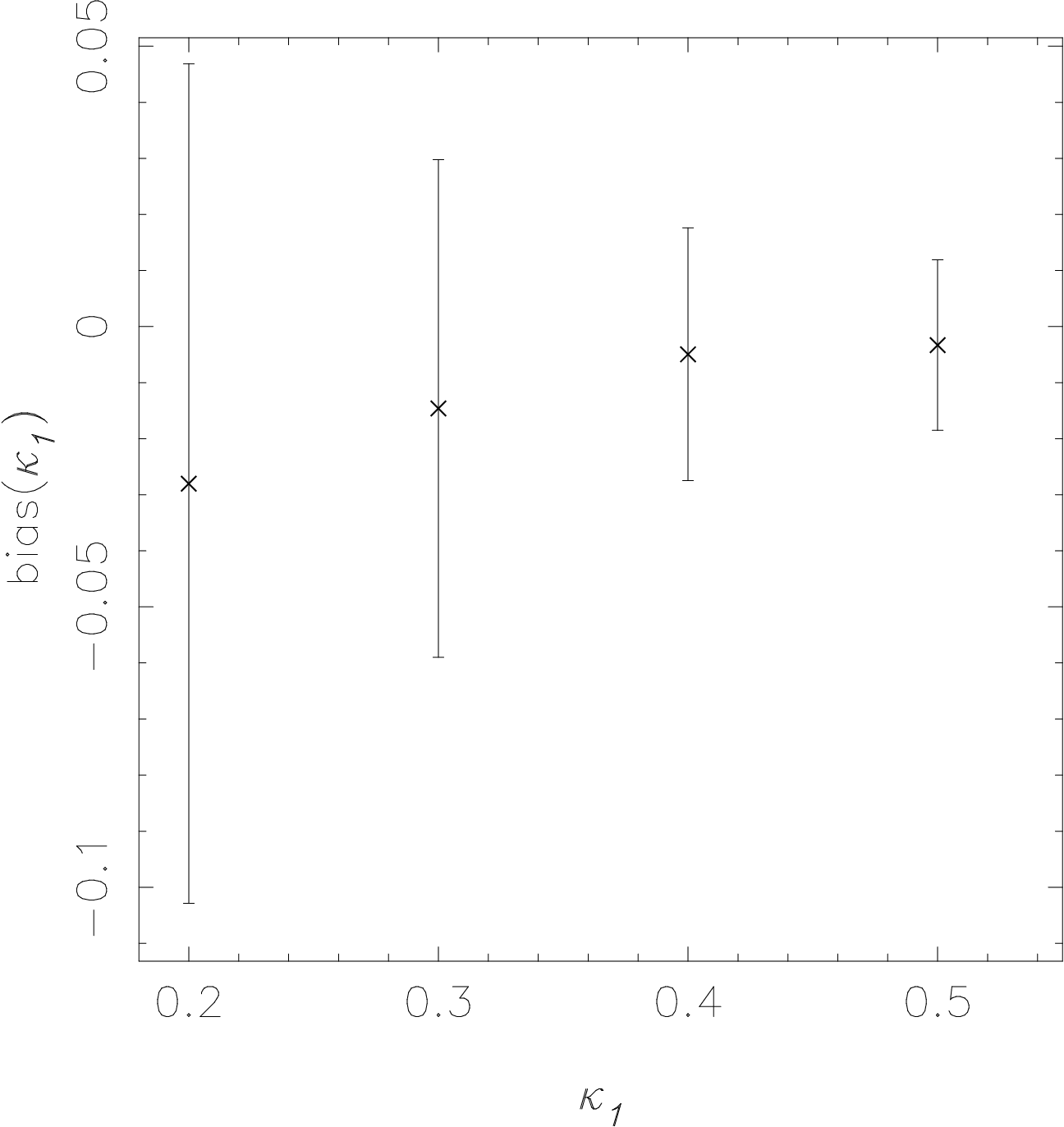}
\centerline{\bf (b)}
\end{center}
\end{minipage}
\end{minipage}
\caption{{\bf (a)} Recovered parameter values (crosses) as a result of
  minimising the log-likelihood function Eq.~\eqref{eq:15}. For each of the
  four sets of parameters (denoted by squares) 100 mock catalogues
  were created . We
  use model Family II and the same family was used to fit the data.
  The model has 2 free parameters ($\kappa_0$ and $\kappa_1$) -- see
  Eq.~\eqref{eq:19}.  Solid and dashed lines correspond to the expected
  mass-sheet degeneracy calculated using $w_i = 1 / \sigma_i$
  and $w_i={\rm const.}$ in Eq.~\eqref{eq:18} for the weighting scheme
  respectively (both almost overlap).  Dot-dashed lines are given by
  $\kappa_0 + \kappa_1 ={\rm const}$. {\bf(b)} Relative bias and variance for
  the recovered values of $\kappa_1$.}
\label{fig:fitlens}
\end{figure*}
Mock catalogues for Family I were generated using $\gamma_1 = \gamma_2 = 0.1$
(see Fig.~\ref{fig:fitconstlens}, panel a) and $\gamma_1 = \gamma_2 = 0.2$
(panel b); 10 different values of $\kappa$ ranging from 0.5 to 1.5
were chosen for both sets.  For each model we fitted 100 mock
catalogues (generated as described above) and we marked with crosses
the resulting best fit parameters (the
original values of the parameters are represented as squares). 
In order to explore the importance of redshift
measurement errors we have generated catalogues without and 
with redshift errors as described above (in  Fig.~\ref{fig:fitconstlens} we
show only the latter).

Surprisingly, even for a relatively strong lens with $\kappa = 0.5$ we are
\textit{not\/} able to break the mass-sheet degeneracy with good
accuracy.  For this model, the relative error on $\kappa$ is of the order
of $20 \%$--$40 \%$, which should be considered very large, given the idealized
conditions used in our simulations.  Hence, the results of our
simulations suggest that it is \textit{very difficult\/} to break the
mass-sheet degeneracy for non-critical lenses using shape information
only.  As expected, for data without redshift errors the fit slightly 
improves, though not substantially.

On the other hand, for critical and close to critical 
lenses, we are {\it effectively able to break the mass-sheet
  degeneracy\/}. The surface mass density is well constrained for
$\kappa \gtrsim 1$. As expected, the results show that galaxies with 
$|g| \simeq 1$ contribute substantially to the removal of the
degeneracy.  If the redshift uncertainty is not added the fit improves
and the constraints are tighter.

Finally, it is interesting to observe that in some cases (e.g. $\kappa
= 1$ and $|\gamma| = 0.14$) we obtain biased results if we use
$p_{\epsilon}$ instead of $p_{\epsilon^\mathrm{err}}$ for calculating
log-likelihood (see Fig.~\ref{fig:fit_pde}). Namely, in
Fig.~\ref{fig:fit_pde}a we use the likelihood function that implies
first convolving the data with $p_{\epsilon^\mathrm{err}}$ and only
then performing lensing transformation [see discussion after
Eq.~\eqref{eq:15}]. This can severely bias results for lenses where
many galaxies have $\abs{g} \simeq 1$. Note that for this case also
the expected mass-sheet degeneracy calculated using $w_i = 1 /
\sigma_i$ in Eq.~\eqref{eq:18} needs to be calculated with
$\sigma_\mathrm{err} = 0$, as we are not correctly accounting for the
measurement errors in this case.

In addition to the global, mass sheet invariance, there is a local
point invariance. Namely, looking at a single image ellipticity we can
not distinguish between $g$ and $1 / g^*$ \citep{seitz95}.  This
invariance is in practice easily broken, either by assuming a profile
or simply by stating that $\kappa$ increases towards the cluster
centre. In the case of all sources being at the same redshift, for the
Family I, however, since $\kappa = \mbox{const.}$ this is a global
invariance. In such a case the log-likelihood function has two
equivalent minima, one corresponding to $g$ and one to $1 /
g^*$. However since we do have the redshift information, this
invariance is broken in most cases.

For our special case with $\kappa = 1$ and $|\gamma| = 0.14$ this is
not so simple. The average absolute value of reduced shear for the
galaxies is $\ave{\abs{g}} \simeq 1$ and therefore for most galaxies
the two minima lie close together. If the probability distribution of
lensed ellipticities is properly accounted for, the resulting
solutions give an unbiased estimate for the true minimum (see
Fig.~\ref{fig:fit_pde}b).  Whereas if this is not the case, as in
Fig.~\ref{fig:fit_pde}a, the results can be biased.

 This special case was presented as an extreme example. In practice
 such a lens is of course unrealistic. However we want to stress that
 critical lenses need to be treated with caution when using
 statistical lensing.

\subsubsection{Family II}
\label{sec:family-ii}
Mock catalogues for Family II were generated using $\kappa_0 = 0.9$,
$\theta_\mathrm{c} = 1.5 \mbox{ arcmin}$, and 4 different values of
$\kappa_1$ ranging from 0.2 to 0.5.  We fitted these data with the
same family of models and using $\kappa_0$ and $\kappa_1$ as free
parameters; note that we fixed the core radius to the same value as
the original profile.  Figure~\ref{fig:fitlens}a shows the results of
log-likelihood minimisation for the data with added redshift errors.
Solid and dashed lines in the figure correspond to the expected
mass-sheet degeneracy calculated using the weighting of $w_i = 1 /
\sigma_i$ and $w_i = \mbox{const.}$ in Eq.~\eqref{eq:18} respectively;
dot-dashed lines give $\kappa_0 + \kappa_1 = \mathrm{const}$.

Finally we plot the bias and the variance of the recovered
values for $\kappa_1$  in
Fig.~\ref{fig:fitlens}b. When the lens becomes critical
the errors are small enough and we are able to constrain $\kappa_1$
with high accuracy. For these lenses we are therefore 
{\it effectively able to break the mass-sheet degeneracy\/}. 
This result is in accordance with the conclusions from the constant-lens model.

Surprisingly, for the case $\kappa_0 = 0.9$ and $\kappa_1 = 0.2$ we do
not see the expected mass-sheet degeneracy, rather, the best-fit
parameters lie along the line $\kappa_0 + \kappa_1 = \mathrm{const}$.
Likely, this is due to the fact that for this model we have galaxies
with $\abs{g} \simeq 1$ close to the centre, where the surface mass
density is $\kappa(0) = \kappa_0 + \kappa_1$.  Since the galaxies with
$\abs{g} \simeq 1$ are the ones that contribute most in breaking the
mass-sheet degeneracy, it is not surprising for this model to see a
degeneracy along the line $\kappa_0 + \kappa_1 = \mathrm{const}$.

If we take $\theta_\mathrm{c}$ to be a free parameter in the fitting,
we obtain the degeneracy $\kappa_0 + \kappa_1 = \mathrm{const}$ for
all four sets of model parameters $\pi_\mathrm{t}$.  Since
$\theta_\mathrm{c}$ is allowed to vary, it can adjust so that the
resulting best-fit model has, for lower values of $\kappa_0$, lower
values of $\theta_\mathrm{c}$.  The region of $\abs{g} \simeq 1$ is
then approximately unchanged as long as $\kappa_0 + \kappa_1 =
\kappa_{0\mathrm{t}} + \kappa_{1\mathrm{t}}$.

\subsection{Ensamble-averaged log-likelihood}
\label{sec:ensamble-aver-log}

Using the log-likelihood function \eqref{eq:15} we can in principle
obtain the best fitting parameters $\pi_\mathrm{max}$ given the
observations and the confidence regions on these parameters.  In order
to obtain the expected errors of parameters obtained from a 
single realization
of data, one can calculate the ensemble-averaged log-likelihood.
Ensemble-averaging also provides a useful test for the behaviour of
log-likelihood function in the asymptotic limit.  Given the redshift
distribution of the background sources $p(z)$ and following
\citet{schneider00} we write down the ensemble-average in the general
form as
\begin{equation}
  \label{eq:21}
  \langle l \rangle(\pi) = n_\mathrm{g} \int \diff^2\theta \, \int \diff^2 \epsilon \, p_{\epsilon}(\epsilon | g)
  \, \int \diff z \, p_z(z) l(\pi) \; ,
\end{equation}
where $n_\mathrm{g}$ gives the number
of galaxies per unit area, and $p_{\epsilon}(\epsilon | g)$ is the
probability distribution of lensed ellipticities calculated using
lens parameters $\pi_{\mathrm t}$.

Asymptotically (i.e.\ when the number of source galaxies
$N_\mathrm{g}$ is very large), the quantity $2 \Delta l$, where
\begin{equation}
  \label{eq:22}
  \Delta l =  \langle l \rangle(\pi) -  \langle l \rangle(\pi_\mathrm{t}) \; ,
\end{equation}
behaves as a random variable following a $\chi^2_{M}$ distribution, where
$M$ is the number of free parameters.  In particular, for Family~I,
where $M = 3$, we expect $68.3\%$, $90\%$, $95.4\%$, and $99\%$  points 
within the levels
$2\,\Delta l = \lbrace 3.53, 6.25, 8.02, 11.2\rbrace$, while for Family~II
($M=2$) the corresponding levels are $2\,\Delta l = \lbrace 2.30, 4.61,
6.17, 9.21\rbrace$.  It is interesting to test the behavior of this
quantity and the accuracy of the asymptotic limit in our case.

Note that, as shown by \citet{geiger98}, the ellipticity distribution
is generally skewed even for relatively small ($\abs{g} = 0.6$) reduced
shears; hence, we need to evaluate all integrations of
Eq.~\eqref{eq:21}, including the one on $\epsilon$, numerically.  The
integration has been carried out using the GNU Scientific library routines.

\begin{figure}[ht!]
\begin{center}
\includegraphics[width=8cm]{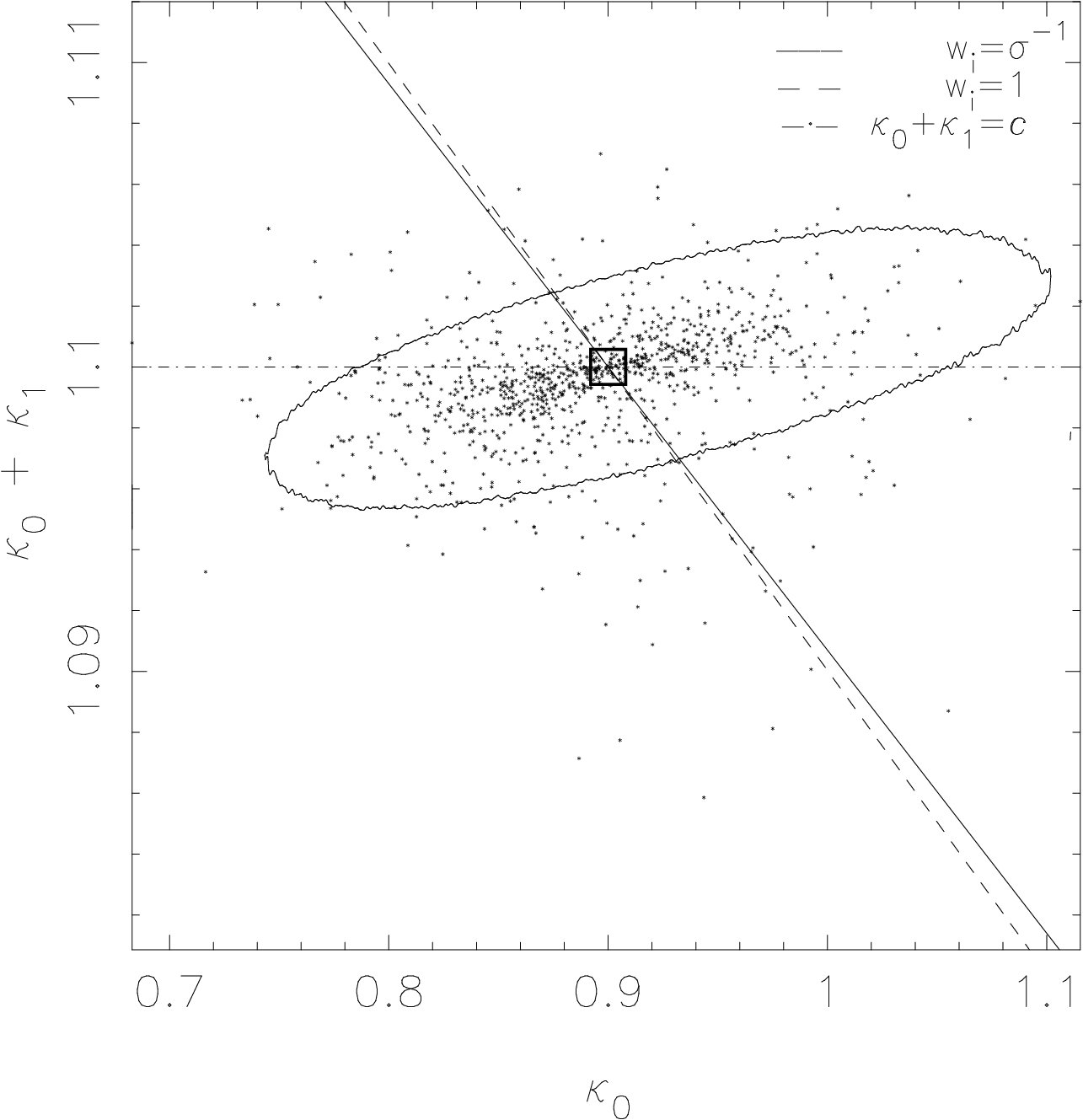}
\end{center}
\caption{Ensemble-averaged log-likelihood for model 
family II with $\kappa_0 = 0.9$ and $\kappa_1 = 0.2$ (square). 
Contour gives the difference of ensemble-averaged 
log-likelihoods with a level of  $2\,\Delta l = 9.21$. In contrast to
previous figures 1000 recovered model parameters are plotted as dots
and $\kappa_0 + \kappa_1$ vs. $\kappa_1$ is plotted (note the small
range on y axis). The 
solid line gives the expected mass-sheet degeneracy line 
calculated with $w_i = 1 / \sigma_i$ in Eq.~\eqref{eq:18}
and dashed line is calculated with $w_i = {\rm const.}$ (both
almost overlapping). Dash-dotted line is given by 
$\kappa_0 + \kappa_1 ={\rm const.}$.} 
\label{fig:ensave}
\end{figure}

In Fig.~\ref{fig:ensave} we show an example of ensemble-average
log-likelihood for one set of simulated parameters for model
Family~II; the other sets give similar results.  We simulated the data
using parameters $\kappa_0 = 0.9$ and $\kappa_1 = 0.2$ without
adding the redshift errors.  This is the extreme case mentioned above,
where the distribution of recovered parameters lies in a very narrow
valley along the line where $\kappa_0 + \kappa_1 = \mathrm{const}$ (dash-dotted
line) rather than along the expected mass-sheet degeneracy line (solid
line and dashed line for different weighting schemes).  The
ensemble-average log-likelihood calculations confirm the anomalous
behavior of the degeneracy. 

The  $2\,\Delta l = 9.21$ contour in
Fig.~\ref{fig:ensave} should enclose $99\%$ of all points. This is not
satisfied in our case; mainly due to the fact that we are not in the
regime of asymptotic limit (seen through the fact that
the points in the plot are not distributed according to Gaussian).

\section{Conclusions}
\label{sec:conclusions}

In this paper we considered a new method to break the mass-sheet
degeneracy in weak lensing mass reconstructions using shape
measurements only.  A detailed analysis of this method has also
clearly shown that breaking the mass-sheet degeneracy is \textit{very
  difficult\/} even in optimal conditions; arguably, it is extremely
difficult in normal observational conditions.

Our main conclusions are summarized in the following items:
\begin{enumerate}
\item The mass-sheet degeneracy can be broken by using
  redshift information of the individual sources.  However, this is
  effective for critical clusters only, i.e.\ for clusters that have
  sizable regions where multiple imaging is possible (and thus perhaps
  observed).  The statistical lensing analysis has to be extended
  close to and inside the critical curves of the cluster.  In the
  regions far outside the critical curve, where weak lensing mass
  reconstructions are normally performed, the lens is too weak for
  the mass-sheet degeneracy to be broken by using redshift and distortion
  information only, even when idealised conditions are employed.
\item Using simulations we find that correlation left-over from the
mass-sheet degeneracy transformation for critical lenses is well
described by Eq.~\eqref{eq:9}, $\kappa \to \kappa' \simeq \lambda
\kappa + (1-\lambda) \bigl\langle Z(z) \bigr\rangle / \bigl\langle
Z^2(z) \bigr\rangle$, where the moments of the cosmological weights
are calculated using $w_i = 1 / \sigma_i$ in \eqref{eq:18}.
\item In order to break the mass-sheet degeneracy with current data it
  is necessary to extend the statistical lensing analysis closer to
  the cluster centre and to simultaneously perform weak and strong
  lensing analysis of the cluster.  This will be a subject of a future
  study.
\end{enumerate}

We are developing a method which combines weak and strong lensing mass
reconstruction techniques simultaneously in a non-parametric fashion.
Weak and strong lensing data has been previously combined by e.g.\
\citet{kneib03}; however, only the combination of weak lensing signal
on scales $>500\mbox{ kpc}$ with the strong lensing signal at $\sim
100 \mbox{ kpc}$ was taken into account. Our approach to breaking the
mass-sheet degeneracy can be employed also for non-parametric mass
reconstruction. The non-parametric statistical lensing reconstruction
technique needs to be extended to and within the critical regions,
resulting in breaking the mass-sheet degeneracy in practice. Such a
method does not rely on any assumption of the parametric form of the
potential.

In summary, although breaking the mass-sheet degeneracy has proven to
be surprisingly difficult, we have shown that it is in principle
possible if one combines constraints on different scales (note that we
assumed the knowledge of the cluster-mass profile, in practice one
obtains the profile by using standard weak-lensing mass
reconstructions). The mass-sheet degeneracy is probably the most
severe limit of current weak lensing mass reconstructions, and is
generally responsible for a significant fraction of the final error on
the ``total'' mass of the cluster.  Hence, breaking the mass-sheet
degeneracy in practice is one of the most important challenges of weak
lensing studies of clusters in the near future.

\begin{acknowledgements}
  We would like to thank Douglas Clowe and Oliver Czoske for many
  useful discussions that helped improve the paper.  This work was
  supported by the International Max Planck Research School for Radio
  and Infrared Astronomy at the University of Bonn, by the Bonn
  International Graduate School, and by the Deutsche
  Forschungsgemeinschaft under the project SCHN 342/3--1.
\end{acknowledgements}

\bibliography{/home/marusa/latex/inputs/bibliogr_clusters,/home/marusa/latex/inputs/bibliogr}

\begin{thebibliography}{28}
\expandafter\ifx\csname natexlab\endcsname\relax\def\natexlab#1{#1}\fi

\bibitem[{{Bartelmann} \& {Narayan}(1995)}]{bartelmann95}
{Bartelmann}, M. \& {Narayan}, R. 1995, \apj, 451, 60

\bibitem[{{Bartelmann} \& {Schneider}(2001)}]{bartelmann00}
{Bartelmann}, M. \& {Schneider}, P. 2001, \physrep, 340, 291

\bibitem[{{Ben{\'{\i}}tez}(2000)}]{benitez00}
{Ben{\'{\i}}tez}, N. 2000, \apj, 536, 571

\bibitem[{{Brainerd} {et~al.}(1996){Brainerd}, {Blandford}, \&
  {Smail}}]{brainerd96}
{Brainerd}, T.~G., {Blandford}, R.~D., \& {Smail}, I. 1996, \apj, 466, 623

\bibitem[{{Bridle} {et~al.}(1998){Bridle}, {Hobson}, {Lasenby}, \&
  {Saunders}}]{bridle98}
{Bridle}, S.~L., {Hobson}, M.~P., {Lasenby}, A.~N., \& {Saunders}, R. 1998,
  \mnras, 299, 895

\bibitem[{{Broadhurst} {et~al.}(1995){Broadhurst}, {Taylor}, \&
  {Peacock}}]{broadhurst95}
{Broadhurst}, T.~J., {Taylor}, A.~N., \& {Peacock}, J.~A. 1995, \apj, 438, 49

\bibitem[{{Clowe} \& {Schneider}(2001)}]{clowe01}
{Clowe}, D. \& {Schneider}, P. 2001, \aap, 379, 384

\bibitem[{{Clowe} \& {Schneider}(2002)}]{clowe02}
---. 2002, \aap, 395, 385

\bibitem[{{Falco} {et~al.}(1985){Falco}, {Gorenstein}, \& {Shapiro}}]{falco85}
{Falco}, E.~E., {Gorenstein}, M.~V., \& {Shapiro}, I.~I. 1985, \apjl, 289, L1

\bibitem[{{Fort} {et~al.}(1997){Fort}, {Mellier}, \& {Dantel-Fort}}]{fort97}
{Fort}, B., {Mellier}, Y., \& {Dantel-Fort}, M. 1997, \aap, 321, 353

\bibitem[{{Geiger} \& {Schneider}(1998)}]{geiger98}
{Geiger}, B. \& {Schneider}, P. 1998, \mnras, 295, 497

\bibitem[{{Hoekstra} {et~al.}(1998){Hoekstra}, {Franx}, {Kuijken}, \&
  {Squires}}]{hoekstra98}
{Hoekstra}, H., {Franx}, M., {Kuijken}, K., \& {Squires}, G. 1998, \apj, 504,
  636

\bibitem[{{James} \& {Roos}(1975)}]{james75}
{James}, F. \& {Roos}, M. 1975, Computer Physics Communications, 10, 343

\bibitem[{{Kaiser}(1995)}]{kaiser95}
{Kaiser}, N. 1995, \apjl, 439, L1

\bibitem[{{Kaiser} \& {Squires}(1993)}]{kaiser93}
{Kaiser}, N. \& {Squires}, G. 1993, \apj, 404, 441

\bibitem[{{Kaiser} {et~al.}(1995){Kaiser}, {Squires}, \& {Broadhurst}}]{ksb95}
{Kaiser}, N., {Squires}, G., \& {Broadhurst}, T. 1995, \apj, 449, 460

\bibitem[{{Kleinheinrich}(2003)}]{kleinheinrich03}
{Kleinheinrich}, M. 2003, {Dark Matter halos of galaxies studied with weak
  gravitational lensing}, PhD Thesis, University of Bonn

\bibitem[{{Kneib}(2003)}]{kneib03}
{Kneib}, J. 2003, \apj, 598, 804

\bibitem[{{Lombardi}(2000)}]{marco_thesis}
{Lombardi}, M. 2000, Ph.D.~Thesis

\bibitem[{{Lombardi} \& {Bertin}(1999)}]{lombardibertin99}
{Lombardi}, M. \& {Bertin}, G. 1999, \aap, 342, 337

\bibitem[{{Marshall} {et~al.}(2002){Marshall}, {Hobson}, {Gull}, \&
  {Bridle}}]{marshall02}
{Marshall}, P.~J., {Hobson}, M.~P., {Gull}, S.~F., \& {Bridle}, S.~L. 2002,
  \mnras, 335, 1037

\bibitem[{{Schneider}(1996)}]{schneider96}
{Schneider}, P. 1996, \mnras, 283, 837

\bibitem[{{Schneider} {et~al.}(1992){Schneider}, {Ehlers}, \& {Falco}}]{sc92}
{Schneider}, P., {Ehlers}, J., \& {Falco}, E. 1992, {Gravitational Lenses}
  (Gravitational Lenses, Springer-Verlag Berlin Heidelberg New York.)

\bibitem[{{Schneider} {et~al.}(2000){Schneider}, {King}, \&
  {Erben}}]{schneider00}
{Schneider}, P., {King}, L., \& {Erben}, T. 2000, \aap, 353, 41

\bibitem[{{Schneider} \& {Seitz}(1995)}]{seitz95}
{Schneider}, P. \& {Seitz}, C. 1995, \aap, 294, 411

\bibitem[{{Seitz} \& {Schneider}(1997)}]{seitz97}
{Seitz}, C. \& {Schneider}, P. 1997, \aap, 318, 687

\bibitem[{{Taylor} {et~al.}(1998){Taylor}, {Dye}, {Broadhurst}, {Benitez}, \&
  {van Kampen}}]{taylor98}
{Taylor}, A.~N., {Dye}, S., {Broadhurst}, T.~J., {Benitez}, N., \& {van
  Kampen}, E. 1998, \apj, 501, 539

\bibitem[{{Tyson} {et~al.}(1990){Tyson}, {Wenk}, \& {Valdes}}]{tyson90}
{Tyson}, J.~A., {Wenk}, R.~A., \& {Valdes}, F. 1990, \apjl, 349, L1

\end{thebibliography}
\bibliographystyle{/home/marusa/latex/inputs/aa}
\end{document}